\documentclass[11pt]{amsart}

\usepackage{latexsym}
\usepackage{amsfonts}
\usepackage{amsthm}
\usepackage{graphicx}
\usepackage{amssymb}
\usepackage{amsmath}
\usepackage{amssymb}
\usepackage{color}
\author[]{Dima Grigoriev}
\address{CNRS, Math\'ematiques, Universit\'e de Lille, 59655, Villeneuve d'Ascq, France}
\email{dmitry.grigoryev@math.univ-lille1.fr}
\thanks{Research of the first author was partially supported by the Federal Agency of the
Science and Innovations of Russia, State Contract No.
02.740.11.5192}

\author[]{Vladimir Shpilrain}
\address{Department of Mathematics, The City  College  of New York, New York,
NY 10031}
\email{shpil@groups.sci.ccny.cuny.edu}
\thanks{Research of the second author was partially supported by
the NSF grants DMS-0914778  and CNS-1117675}

\newtheorem{problem}{Problem}
\newtheorem{example}{Example}

\newtheorem{proposition}{Proposition}
\newtheorem{remark}{Remark}

\begin{document}

\title[Secrecy without one-way functions]{Secrecy without one-way functions}

\begin{abstract}
We show that some problems in information security can be solved
without using one-way functions. The latter are usually regarded as
a central concept of cryptography, but the very existence of one-way
functions depends on difficult conjectures in complexity theory,
most notably on the notorious   ``$P \ne NP$"   conjecture. This is
why cryptographic primitives that do not employ one-way functions
are often called  ``unconditionally secure".

In this paper, we suggest  protocols for secure computation of the
sum, product, and some other functions of two or  more elements of
an arbitrary constructible ring, without using any one-way
functions. A new input that we offer here is that, in contrast with
other proposals, we conceal ``intermediate results" of a
computation. For example, when we compute the sum of $k$ numbers,
only the final result is known to the parties; partial sums are not
known to anybody. Other applications of our method include
voting/rating over insecure channels and a rather elegant and
efficient  solution of the ``two millionaires problem".

Then, while it is fairly obvious that a secure (bit) commitment
between two parties is impossible without a one-way function, we
show that it is possible if the number of parties is at least 3. We
also show how our unconditionally secure (bit) commitment scheme for
3 parties can be used to arrange an unconditionally secure (bit)
commitment between just two parties if they use a ``dummy" (e.g., a
computer) as the third party. We explain how our concept  of a
``dummy" is different from a well-known concept  of a ``trusted
third party". Based on a similar idea, we also offer an
unconditionally secure $k$-$n$ oblivious transfer protocol between
two parties who use a ``dummy".

We also suggest a protocol, without using a one-way function, for
the so-called ``mental poker", i.e., a fair card dealing (and
playing) over distance.

Finally,  we propose a secret sharing scheme where an   advantage
over Shamir's  and other known secret sharing schemes is that
nobody, including the dealer, ends up knowing  the shares (of the
secret) owned by any particular player.

It should be mentioned that  computational cost of our protocols is
negligible to the point that all of them can be  executed without a
computer.

\end{abstract}

\maketitle

\section{Introduction}

Secure multi-party computation  is a problem that was originally
suggested by Yao \cite{Yao} in  1982. The concept usually refers to
computational systems in which several parties wish to jointly
compute some value based on individually held secret bits of
information, but do not wish to reveal their secrets to   anybody in
the process. For example, two individuals who each possess some
secret numbers, $x$ and $y$, respectively, may wish to jointly
compute some function $f(x,y)$ without revealing any information
about $x$ or $y$ other than what can be reasonably deduced by
knowing the actual value of $f(x,y)$.

Secure computation was formally introduced  by  Yao as secure
two-party computation. His ``two millionaires problem" (cf. our
Section \ref{negative}) and its solution gave way to a
generalization to multi-party protocols, see e.g. \cite{Chaum},
\cite{DI}. Secure multi-party computation provides solutions to
various real-life problems such as distributed voting, private
bidding and auctions, sharing of signature or decryption functions,
private information retrieval, etc.

In this paper, we offer protocols for secure computation of the sum
and  product of three or  more elements of an arbitrary
constructible ring without using encryption or any one-way functions
whatsoever. We require in our scheme that there are $k$ secure
channels for communication between the $k \ge 3$ parties, arranged
in a circuit. We also show that less than $k$ secure channels is not
enough.

Unconditionally secure  multiparty computation was previously
considered in \cite{Chaum} and elsewhere ({\it since the present
paper is not a survey, we do not give a comprehensive bibliography
on the subject here, but only mention what is most relevant to our
paper}). A new input that we offer here is that, in contrast with
\cite{Chaum} and other proposals, we conceal ``intermediate results"
of a computation. For example, when we compute a sum of $k$ numbers
$n_i$, only the final result ~$\sum_{i=1}^k n_i$ is known to the
parties; partial sums are not known to anybody. This is not the case
in \cite{Chaum} where each   partial sum ~$\sum_{i=1}^s n_i$ is
known to at least some of the parties. This difference is important
because, by the ``pigeonhole principle", at least  one of the
parties may accumulate sufficiently many  expressions in $n_i$ to be
able  to recover at least some of the $n_i$ other than his own.

Here we show how our method works for computing the sum (Section
\ref{sum}) and the product (Section \ref{product}) of private
numbers. We ask what other functions can be securely computed
without revealing intermediate results.

Other applications of our method include voting/rating over insecure
channels (Section \ref{voting}) and a rather elegant solution of the
``two millionaires problem" (Section \ref{negative}).


We also address  another cryptographic primitive, known as (bit)
commitment. In cryptography, a commitment scheme  allows one to
commit to a value while keeping it hidden, with the ability to
reveal the committed value later. Commitments are used to bind a
party to a value so that they cannot adapt to other messages in
order to gain some kind of inappropriate advantage. They are
important to a variety of cryptographic protocols including secure
coin flipping, zero-knowledge proofs, and secure multi-party
computation. See \cite{OG} or \cite{Menezes}  for a general
overview.

It is known \cite{Luby} that a secure (bit) commitment between two
parties is impossible without some kind of encryption, i.e., without
a one-way function. However, if the number of parties is at least 3,
this becomes possible, as long as parties do not form coalitions to
trick other party (or parties). It has to be pointed out though that
formal definitions of commitment schemes vary strongly in notation
and in flavor, so we have to be specific about our model. We give
more formal details in Section \ref{commitment}, while here we just
say, informally, that what we achieve is the following: if the
committed  values are just bits, then after the commitment stage of
our scheme is completed, none of the parties can guess any other
party's bit with probability greater than $\frac{1}{2}$. We require
in our scheme that there are $k$ secure channels for communication
between the  parties, arranged in a circuit. We also show that less
than $k$ secure channels is not enough.

Then, in Section \ref{commitment_2}, we show how our unconditionally
secure (bit) commitment scheme for 3 parties can be used to arrange
an unconditionally secure (bit) commitment between just two parties
if they use a ``dummy" (e.g., a computer) as the third party. We
explain how our concept  of a ``dummy" is different from a
well-known concept  of a ``trusted third party"  and  also from
Rivest's idea of a ``trusted initializer" \cite{Rivest}. In
particular, a  difference important for real-life applications  is
that our ``dummy" is unaware of the  committed values. Also, our
``dummy" is {\it passive}, i.e., he does not privately transmit
information to ``real" participants and he does not generate
randomness.

Based on a similar idea, we also offer, in Section \ref{oblivious},
an unconditionally secure $k$-$n$ oblivious transfer protocol
between two parties who use a ``dummy".

In Section \ref{poker}, we consider a related cryptographic
primitive known as ``mental poker", i.e., a fair card dealing (and
playing) over distance. Several protocols for doing this, most of
them using encryption, have been suggested, the first by Shamir,
Rivest, and Adleman \cite{SRA}, and subsequent proposals include
\cite{Crepeau} and \cite{GM}. As with the bit commitment,  a fair
card dealing between just two players over distance is impossible
without a one-way function since commitment is part of any
meaningful card dealing scenario. However, it turns out to be
possible if the number of players is $k \ge 3$. What we require
though is that there are $k$ secure channels for communication
between players, arranged in a circuit.  We also show that our
protocol can, in fact, be adapted to deal cards to just 2 players.
Namely, if we have 2 players, they can use a ``dummy" player (e.g. a
computer), deal cards to 3 players, and then just ignore the
``dummy"'s cards, i.e., ``put his cards back in the deck".  An
assumption on the ``dummy" player is that he cannot generate any
randomness, so randomness has to be supplied to him by the two
``real" players. Another assumption is that there are secure
channels for communication between either ``real" player and the
``dummy". We believe that this model is adequate for 2 players who
want to play online but do not trust the server. ``Not trusting" the
server exactly means not trusting with generating randomness. Other,
deterministic, operations can be verified at the end of the game; we
give more details in Section \ref{Protocol2}.

We note that the only known (to us) proposal for dealing cards to $k
\ge 3$ players over distance without using one-way functions was
published in \cite{BF}, but their protocol lacks the simplicity,
efficiency, and some of the functionalities of our proposal; this is
discussed in more detail in our Section \ref{properties}. Here we
just mention that computational cost of our protocols is negligible
to the point that  they can be easily executed  without a computer.

Finally, in Section \ref{secret}, we propose a secret sharing scheme
where an   advantage over Shamir's \cite{Shamir} and other known
secret sharing schemes is that nobody, including the dealer, ends up
knowing  the shares (of the secret) owned by any particular players.
The disadvantage though is that our scheme is a $(k, k)$-threshold
scheme only.


\section{Secure computation of a sum}
\label{sum}

In this section, our scenario is as follows. There are $k$ parties
$P_1, \ldots, P_k$; each $P_i$ has a private element $n_i$ of a
fixed constructible ring $R$. The goal is to compute the sum of all
$n_i$ without revealing any of the $n_i$  to any party $P_j, j \ne
i$.


One obvious way to achieve this is well studied in the literature
(see e.g. \cite{OG, GM, Grigoriev2}): encrypt each $n_i$ as
$E(n_i)$, send all $E(n_i)$ to some designated $P_i$ (who does not
have a decryption key), have $P_i$ compute $S = \sum_i E(n_i)$ and
send the result to the participants for decryption. Assuming that
the encryption function $E$ is {\it homomorphic}, i.e., that $\sum_i
E(n_i) = E(\sum_i n_i)$, each party $P_i$ can recover $\sum_i n_i$
upon decrypting $S$.

This scheme requires not just a one-way function, but a one-way
function with a trapdoor since both encryption and decryption are
necessary to obtain the result.


What we suggest in this section is a protocol that does not require
any one-way function, but involves secure communication between some
of the $P_i$. So, {\it our assumption} here is that there are $k$
secure channels of communication between  the $k$ parties $P_i$,
arranged in a circuit. {\it Our result} is computing the sum of
private elements $n_i$ without revealing any individual $n_i$ to any
$P_j, j \ne i$. Clearly, this is only possible if the number of
participants $P_i$ is greater than 2. As for the number of secure
channels between $P_i$, we will  show that  it cannot be less than
$k$, by the number of parties.

\subsection{The protocol (computing the sum)}

\begin{enumerate}

\item $P_1$ initiates the process by sending $n_1+n_{01}$ to $P_2$, where $n_{01}$ is a random
element (``noise").

\item Each $P_i, ~2 \le i \le k-1$,   does the following. Upon
receiving an element $m$ from $P_{i-1}$, he adds his $n_i+n_{0i}$ to
$m$ (where $n_{0i}$ is a random element) and sends the result to
$P_{i+1}$.

\item   $P_k$ adds   $n_k+n_{0k}$ to whatever he has received from
$P_{k-1}$ and sends the result to $P_{1}$.

\item $P_{1}$ subtracts $n_{01}$ from what he got from $P_k$; the
result now is the sum  $S = \sum_{1 \le i \le k} n_i ~+ ~\sum_{2 \le
i \le k} n_{0i}$. Then  $P_{1}$ publishes $S$.

\vskip .2 cm

\item Now all participants $P_i$, except $P_1$, broadcast their $n_{0i}$, possibly
over insecure channels, and compute $\sum_{2 \le i \le k} n_{0i}$.
Then they subtract the result from $S$ to finally get $\sum_{1 \le i
\le k} n_i$.

%
%

\end{enumerate}

Thus, in this protocol we have used $k$ (by the number of the
parties $P_i$) secure channels of communication between the parties.
If we visualize the arrangement as a graph with $k$ vertices
corresponding to the parties $P_i$ and $k$ edges corresponding to
secure channels, then this graph will be a $k$-cycle. Other
arrangements are possible, too; in particular, a union of disjoint
cycles of length $\ge 3$ would do. (In that case, the graph will
still have $k$ edges.) Two natural questions that one might now ask
 are: (1) is  any
arrangement with less than $k$ secure channels possible?  (2) with
$k$ secure channels, would this scheme work with any arrangement
other than a union of disjoint cycles of length $\ge 3$? The answer
to both questions is ``no". Indeed, if there is a vertex
(corresponding to $P_1$, say) of degree 0, then any information sent
out by $P_1$ will be available to everybody, so other participants
will know $n_1$ unless $P_1$ uses a one-way function to conceal it.
If there is a vertex (again, corresponding to $P_1$) of degree 1,
this would mean that $P_1$ has a secure channel of communication
with just one other participant, say $P_2$. Then any information
sent out by $P_1$ will be available at least to $P_2$, so $P_2$ will
know $n_1$ unless $P_1$ uses a one-way function to conceal it. Thus,
every vertex in the graph should have degree at least 2, which
implies that every vertex is included in a cycle. This immediately
implies that the total number of edges is at least $k$. If now a
graph $\Gamma$ has $k$ vertices and $k$ edges,  and every vertex of
$\Gamma$ is included in a cycle, then every vertex has degree
exactly 2 since by the ``handshaking lemma" the sum of the  degrees
of all vertices in any graph equals twice the number of edges. It
follows that our graph is a union of disjoint cycles.

\subsection{Effect of coalitions}
\label{coalitions}

Suppose now we have $k \ge 3$ parties with $k$ secure channels of
communication arranged in a circuit, and suppose 2 of the parties
secretly form a coalition. Our assumption here is that, because of
the circular arrangement of  secure channels, a secret coalition is
only possible between parties $P_i$ and $P_{i+1}$ for some $i$,
where the indices are considered modulo $k$; otherwise, attempts to
form a coalition (over insecure channels) will be detected. If two
parties $P_i$ and $P_{i+1}$ exchanged information, they would, of
course, know each other's elements $n_i$, but other than that, they
would not get any advantage if  $k \ge  4$. Indeed, we can just
``glue these two parties together", i.e., consider them as one
party, and then the protocol is essentially reduced to that with
$k-1 \ge 3$ parties. On the other hand, if $k=3$, then, of course,
two parties together have all the information about the third
party's element.

For an arbitrary $k \ge  4$, if $n < k$ parties want to form a
(secret) coalition to get information about some other party's
element, all these $n$ parties have to be connected by secure
channels, which means there is a $j$ such that these $n$ parties are
$P_j, P_{j+1}, \ldots, P_{j+n-1}$, where indices are considered
modulo $k$. It is not hard to see then that only a coalition of
$k-1$ parties $P_1, \ldots, P_{i-1}, P_{i+1}, \ldots, P_k$ can
suffice to get information about the $P_i$'s element.

\subsection{Ramification: voting/rating over insecure channels}
\label{voting}

In this section, our scenario is as follows. There are $k$ parties
$P_1, \ldots, P_k$; each $P_i$ has a private integer $n_i$. There is
also a computing entity $B$ (for Boss) who shall compute the sum of
all $n_i$. The goal is to let $B$ compute the sum of all $n_i$
without revealing any of the $n_i$ to him or to any party $P_j, j
\ne i$.

The following example from real life is a motivation for this
scenario.

\begin{example}
Suppose members of the board in a company have to vote for a project
by submitting their numeric scores (say, from 1 to 10) to the
president of the company. The project gets a green light if the
total score is above some threshold  value $T$. Members of the board
can discuss the project between themselves and exchange information
 privately, but none of them  wants his/her score to be known to
either the president or any other member of the board.
\end{example}

In the protocol below, we are again assuming that there are $k$
  channels of communication between the parties, arranged in a
circuit: $P_1 \to P_2 \to \ldots \to P_k \to  P_1$. On the other
hand, communication channels between B and any of the parties are
not assumed to be secure.

\subsection{The protocol (rating over insecure channels) }

\begin{enumerate}

\item $P_1$ initiates the process by sending $n_1+n_{01}$ to $P_2$, where $n_{01}$ is a random
number.

\item Each $P_i, ~2 \le i \le k-1$,   does the following. Upon
receiving a number $m$ from $P_{i-1}$, he adds his $n_i+n_{0i}$ to
$m$ (where $n_{0i}$ is a random number) and sends the result to
$P_{i+1}$.

\item   $P_k$ adds   $n_k+n_{0k}$ to whatever he has received from
$P_{k-1}$ and sends the result to B.

\item $P_{k}$ now starts the process of collecting the ``adjustment" in the
opposite direction. To that effect, he sends his $n_{0k}$ to
$P_{k-1}$.

\item $P_{k-1}$ adds $n_{0(k-1)}$ and sends the result to
$P_{k-2}$.

\item The process ends when $P_1$ gets a number from $P_2$, adds
his $n_{01}$,  and sends the result to B. This  result is the sum of
all $n_{0i}$.

\item B subtracts what he got from $P_1$ from what he got from $P_k$; the
result now is the sum of all $n_i, ~1 \le i \le k$.

\end{enumerate}

\section{Application: the ``two millionaires problem"}
\label{negative}

The  protocol from Section \ref{sum}, with some adjustments, can be
used to provide  an elegant and efficient solution to the ``two
millionaires problem" introduced in \cite{Yao}: there are two
numbers,  $n_1$ and $n_2$, and the goal is to solve the inequality
$n_1 \geq n_2 ?$ without revealing the actual values of $n_1$ or
$n_2$.

To that effect, we use a ``dummy"  as the third party. Our concept
of a ``dummy" is quite different from a well-known concept  of a
``trusted third party"; importantly, our ``dummy" is not supposed to
generate any randomness; he just does what he is told to. Basically,
the only difference between our ``dummy" and a usual calculator is
that there are secure channels of communication between the ``dummy"
and either ``real" party. One possible real-life interpretation of
such a ``dummy" would be an online calculator that can combine
inputs from different users. Also note that in our scheme below {\it
the ``dummy" is unaware of the committed values} of $n_1$ or $n_2$,
which is useful in case the two ``real" parties do not want their
private numbers to ever be revealed. This suggests yet another
real-life interpretation of a ``dummy", where he is a mediator
between two parties negotiating a settlement.


Thus, let A (Alice) and B  (Bob) be two ``real" parties, and D
(Dummy) the ``dummy". Suppose A's number is $n_1$, and  B's number
is $n_2$.

\subsection{The protocol (comparing two numbers)}

\begin{enumerate}

\item  A splits her number $n_1$ as a difference  $n_1=n_1^+ -
n_1^-$. She then sends $n_1^-$ to B.

\item  B splits his number $n_2$ as a difference  $n_2=n_2^+ -
n_2^-$. He then sends $n_2^-$ to A.

\item  A sends $n_1^+ + n_2^-$ to D.

\item  B sends $n_2^+ + n_1^-$ to D.

\item D subtracts $(n_2^+ + n_1^-)$ from $(n_1^+ + n_2^-)$ to get
$n_1-n_2$, and   announces whether this result is positive or
negative.

\end{enumerate}

%
%
%
%
%
%


\begin{remark}
Perhaps a point of some dissatisfaction in this protocol could be
the fact that the ``dummy" ends up knowing the actual difference
$n_1-n_2$, so if there is a leak of this information to either
party, this party would recover the other's private number $n_i$.
This can be avoided if $n_1$  and $n_2$ are represented in the
binary form and compared one bit at a time, going left to right,
until the difference between bits becomes nonzero. However, this
method, too, has a disadvantage: the very moment the ``dummy"
pronounces the difference between bits nonzero would give an {\em
estimate} of the difference $n_1-n_2$ {\em to the real parties}, not
just to the ``dummy".
\end{remark}

%
%
%
%
%
%
%
%
%

We note that the original solution of the ``two millionaires
problem" given in \cite{Yao}, although lacks the elegance of our
scheme, does not involve a third party, whereas our solution does.
On the other hand, the solution in \cite{Yao} uses encryption,
whereas our solution does not, which makes it by far more efficient.

\section{Secure computation of a product}
\label{product}

In this section, we show how to use the same general ideas from
Section \ref{sum} to securely compute a product. Again, there are
$k$ parties $P_1, \ldots, P_k$; each $P_i$ has a private (nonzero)
element $n_i$ of a fixed constructible ring $R$. The goal is to
compute the product of all $n_i$ without revealing any of the $n_i$
to any party $P_j, j \ne i$. Requirements on the ring $R$ are going
to be somewhat more stringent here than they were in Section
\ref{sum}. Namely, we require that $R$ does not have zero divisors
and, if an element $r$ of $R$ is a product $a \cdot x$ with a known
$a$ and an unknown $x$, then  $x$ can be efficiently recovered from
$a$ and $r$. Examples of rings with these properties include the
ring of integers and any constructible field.

\subsection{The protocol (computing the product)}

\begin{enumerate}

\item   $P_1$ initiates the process by sending $n_1 \cdot n_{01}$ to $P_2$, where $n_{01}$ is a
random nonzero element (``noise").

\item Each $P_i, ~2 \le i \le k-1$,   does the following. Upon
receiving an element $m$ from $P_{i-1}$, he multiplies $m$ by $n_i
\cdot n_{0i}$ (where $n_{0i}$ is a random element)  and sends the
result to $P_{i+1}$.

\item   $P_k$ multiplies by $n_k \cdot n_{0k}$   whatever he has received from
$P_{k-1}$ and sends the result to $P_{1}$. This  result is the
product $P = \Pi_{1 \le i \le k} ~n_i ~\cdot ~\Pi_{2 \le i \le k}
~n_{0i}$.

\item $P_{1}$   divides what he got from $P_k$ by his $n_{01}$; the
result now is the product $P = \Pi_{1 \le i \le k} ~n_i ~\cdot
~\Pi_{2 \le i \le k} ~n_{0i}$. Then  $P_{1}$ publishes $P$.

\item Now all participants $P_i$, except $P_1$, broadcast their $n_{0i}$, possibly
over insecure channels, and compute $\Pi_{2 \le i \le k} ~n_{0i}$.
Then they  divide $P$ by the result to finally get $\Pi_{1 \le i \le
k} ~n_i$.


\end{enumerate}

\section{Secure computation of symmetric functions}
\label{symmetric}

In this section, we show how our method can be easily generalized to
allow secure computation of any expression of the form
~$\sum_{i=1}^k n_i^r$, where $n_i$ are parties' private numbers, $k$
is the number of parties, and $r \ge 1$ an arbitrary integer. We
simplify our method here by removing the ``noise", to make the
exposition more transparent.

\subsection{The protocol (computing the sum of powers)}

\begin{enumerate}

\item   $P_1$ initiates the process by sending a random element
$n_0$ to $P_2$.

\item Each $P_i, ~2 \le i \le k-1$,   does the following. Upon
receiving an element $m$ from $P_{i-1}$, he adds his $n_i^r$ to $m$
and sends the result to $P_{i+1}$.

\item   $P_k$ adds his $n_k^r$ to whatever he has received from
$P_{k-1}$ and sends the result to $P_{1}$.

\item $P_{1}$ subtracts $(n_0-n_1^r)$ from what he got from $P_k$; the
result now is the sum of all $n_i^r, ~1 \le i \le k$.

\end{enumerate}

Now that the parties can securely compute the sum of any powers of
their $n_i$, they can also compute any symmetric function of $n_i$.
However, in the course of computing a symmetric function from sums
of different powers of $n_i$, at least some of the parties will
possess several different polynomials in $n_i$, so chances are that
at least some of the parties will be able to recover at least some
of the $n_i$. On the other hand, because of the symmetry of all
expressions involved, there is no way to tell which $n_i$ belongs to
which party.

\subsection{Open problem}

Now it is natural to ask:

\begin{problem}
What other functions (other than the sum and the product) can be
securely computed without revealing intermediate results to any
party?
\end{problem}

To be more precise, we note that one intermediate result is
inevitably revealed to the party who finishes computation, but this
cannot be avoided in any scenario. For example, after the parties
have computed the sum of their private numbers, each  party also
knows the sum of all numbers except his own. What we want is that no
other intermediate results are ever revealed.

To give some insight into this problem, we consider a couple of
examples of computing  simple functions different from  the sum and
the product of the parties' private numbers.

\begin{example}
We show how to compute the function $f(n_1, n_2, n_3) = n_1 n_2 +
n_2 n_3$ in the spirit of the present paper, without revealing (or
even computing) any intermediate results, i.e., without  computing
$n_1 n_2$  or  $n_2 n_3$.

\begin{enumerate}

\item   $P_2$ initiates the process by sending a random element
$n_0$ to $P_3$.


\item $P_3$ adds his  $n_3$ to $n_0$ and sends $n_3+n_0$  to
$P_1$.

\item   $P_1$ adds his $n_1$ to $n_0+ n_3$ and sends the result to
$P_2$.

\item $P_2$ subtracts $n_0$ from $n_0 + n_3 + n_1$ and
multiplies the result by $n_2$. This is now $n_1 n_2 + n_2 n_3$.

\end{enumerate}

\end{example}

\begin{example}
The point of this example is to show that functions that can be
computed by our method do not have to be homogeneous (in case the
reader got this impression based on the previous examples).

The function that we compute here is $f(n_1, n_2, n_3) = n_1 n_2 +
g(n_3)$, where $g$ is any computable function.

\begin{enumerate}

\item   $P_1$ initiates the process by sending a random element
$a_0$ to $P_2$.

\item $P_2$ multiplies $a_0$ by his  $n_2$ and sends the result to
$P_3$.

\item $P_3$ multiplies $a_0 n_2$ by a random element $c_0$ and sends the result
to $P_1$.

\item   $P_1$  multiplies $a_0 n_2 c_0$ by his $n_1$, divides by
$a_0$, and sends the result, which is $n_1 n_2 c_0$, back to $P_3$.

\item $P_3$ divides $n_1 n_2 c_0$ by $c_0$ and adds $g(n_3)$, to end
up with $n_1 n_2 + g(n_3)$.

\end{enumerate}

\end{example}

Note that in this example, the parties used more than just one loop
of transmissions in the course of computation. Also, information
here was sent ``in both directions" in the circuit.

\begin{remark}
Another collection of examples of multiparty computation  without
revealing  intermediate results can be obtained as follows. Suppose,
without loss of generality, that some function $f(n_1, \ldots, n_k)$
can be computed by our method in such a way that the last step in
the computation is performed by the party $P_1$, i.e., $P_1$ is the
one who ends up with $f(n_1, \ldots, n_k)$ while no
 party knows any intermediate  result $g(n_1, \ldots, n_k)$ of this computation.
Then, obviously, $P_1$ can produce any function of the form $F(n_1,
f(n_1, \ldots, n_k))$ (for a computable function $F$) as well.
Examples include $n_1^r + n_1 n_2 \cdots n_k$ for any $r \ge 0$;
$n_1^r + (n_1 n_2 + n_3)^s$ for any $r, s \ge 0$, etc., etc.

\end{remark}

\section{(Bit) commitment}
\label{commitment}

While it is fairly obvious that a secure (bit) commitment between
two parties is impossible without a one-way function, we show here
that it is possible if the number of parties is at least 3.
Generalizing the standard concept (see e.g. \cite{OG}) of a
two-party (bit) commitment scheme, we define an $n$-party (bit)
commitment scheme to be a two-phase protocol through which each of
the  $n$ parties  can commit himself to a value such that the
following two requirements are satisfied:
\medskip

\noindent {\bf (1)} Secrecy: at the end of the {\it commitment}
phase, none of the $n$ parties gains any information about any other
party's committed value.
\medskip

\noindent {\bf (2)} Unambiguity: suppose that the commitment phase
is successfully completed. Then, if later the parties perform the
{\it decommitment} phase (sometimes called the {\it reveal} phase),
each party's  committed value can be recovered (collectively by
other parties) without ambiguity.
\medskip

To make our ideas more transparent, we start with the simplest case
where there are just 3 parties: $P_1$, $P_2$, and $P_3$, and no two
of them form a coalition against the third  one. Suppose they want
to commit to   integers $n_1$, $n_2$, and $n_3$ (modulo some $m \ge
2$), respectively. More precisely, the scenario is as follows.
During the commitment phase, the parties exchange various pieces of
information about their integers $n_i$. After that, the parties
``decommit", or reveal, their integers and prove to each  other that
the integers $n_i$ that they revealed are the same that they
committed to.

All computations below are performed modulo a fixed integer $m \ge
2$.

%



\begin{enumerate}

\item Each  participant $P_i$ randomly splits his integer $n_i$ in a
sum of two  integers: $n_i=r_i+s_i$. If the participants want to
commit to bits rather  than integers, then $P_i$ would split the
``0" bit as either 0+0 or 1+1, and the ``1" bit as either 0+1 or
1+0.

\item ({\it Commitment phase.}) $P_1$ sends $r_1$ to $P_2$, ~then $P_2$ sends $r_1 + r_2$ to
$P_3$, ~then $P_3$ sends $r_1 + r_2 + r_3$ to $P_1$. In the
``opposite direction", $P_3$ sends $s_3$ to $P_2$, ~then $P_2$ sends
$s_2 + s_3$ to $P_1$, ~then $P_1$ sends $s_1 + s_2 + s_3$ to $P_3$.

\begin{quote}
After the commitment phase, $P_1$ has $s_1$, $s_2 + s_3$, $r_1$, and
$r_1 + r_2 + r_3$ (therefore also $r_2 + r_3$), so he cannot
possibly recover any $n_i$ other than his own. (He can recover
$n_2+n_3$, but this does not give him any information about either
$n_2$ or $n_3$). Then, $P_2$ has $s_2$, $s_3$, $r_1$, and $r_2$, so
he, too, cannot possibly recover any $n_i$ other  than his own.
Finally, $P_3$ has $s_3$, $r_3$, $r_1 + r_2$, and $s_1 + s_2 + s_3$
(therefore also $s_1 + s_2$), so he, too, cannot possibly recover
any $n_i$ other than his own, (He can recover $n_1+n_2$, but this
does not give him any information about either $n_1$ or $n_2$).

\end{quote}

\item ({\it Decommitment phase starts.}) Note that during the decommitment
steps below, each   participant transmits information that {\it
somebody else} had committed to before. This way, each piece of
transmitted information can be corroborated by two parties, which
prevents cheating since we are assuming that  no two participants
form a coalition.


\item $P_3$  sends $n_1+n_2$ to both $P_1$  and $P_2$.  Now  $P_1$
knows $n_2$, and $P_2$  knows $n_1$.

\item $P_2$ sends $r_1$ to $P_3$. Now $P_3$ can recover $r_2$ from
$r_1$ and $r_1 + r_2$.

\item $P_1$ sends $s_2 + s_3$ to  $P_3$. Now $P_3$ can extract
$s_2$ from this sum, and then, since he has $r_2$, recover $n_2$,
and then also  $n_1$ since $P_3$ already knows $n_1+n_2$.

\item $P_1$ sends $r_1 + r_2 + r_3$ to  $P_2$. Now $P_2$ can recover
$r_3$ and therefore $n_3=r_3+s_3$.


\end{enumerate}

This protocol can be obviously generalized to $3m$ participants for
arbitrary $m \ge 1$ by splitting the players into triples and
applying the above protocol to each   triple. It can also be
generalized to an arbitrary number $k \ge 3$ of participants with a
circular arrangement of $k$ secure channels, but we leave details to
the reader.

\begin{remark}
A question that one might now ask, if only out of curiosity, is:
would this scheme work with any arrangement of secure channels other
than a union of disjoint circuits of length $\ge 3$? The answer to
this question is ``no". Indeed, if in the graph of secure channels
there is a vertex (corresponding to $P_1$, say) of degree 0, then
any information sent out by $P_1$ will be available to everybody, so
other participants will know $n_1$ unless $P_1$ uses a one-way
function to conceal it. If there is a  vertex (again, corresponding
to $P_1$) of degree 1, this would mean that $P_1$ has a secure
channel of communication with just one other participant, say $P_2$.
Then any information sent out by $P_1$ will be available at least to
$P_2$, so $P_2$ will know $n_1$ unless $P_1$ uses a one-way function
to conceal it. So, every vertex in the graph should have degree at
least 2, which implies that every vertex is included in a circuit.
It follows, in particular, that the total number of secure channels
should be at least $k$, by the number of participants.

\end{remark}

\section{(Bit) commitment between two parties}
\label{commitment_2}

Now we show how our unconditionally secure  commitment scheme for 3
parties from Section \ref{commitment} can be used to arrange an
unconditionally secure  commitment between just two parties. This is
similar, in spirit, to the idea of Rivest  \cite{Rivest}, where an
extra participant is introduced to bring the number of parties up to
3. However, an important difference between our proposal and that of
\cite{Rivest} is that the extra participant in \cite{Rivest} is a
``trusted initializer", which means that
{\bf (i)} he is allowed to generate randomness;   {\bf (ii)} he can
{\it  transmit} information {\it to} ``real" participants over
secure channels.

By contrast, our extra participant is a ``dummy", i.e.,
{\bf (i)} he is not allowed to generate randomness; {\bf (ii)}  he
can  {\it receive} information {\it from} ``real" participants over
secure channels and perform simple arithmetic operations.

One possible real-life interpretation of such a ``dummy" would be an
online calculator that can combine inputs from different users. Also
note that in our scheme below {\it  the ``dummy" is unaware of the
committed values}, which is useful in case the two ``real"
participants do not want their commitments to ever be revealed to
the third party; for example, such a  ``dummy" could be a mediator
between two parties negotiating a divorce settlement.


Thus, let A (Alice) and B  (Bob) be two ``real" participants, and D
(Dummy) the ``dummy". Suppose A   and B want to commit to  integers
$n_1$ and $n_2$, respectively.

\begin{enumerate}

\item  A   and B randomly split their  integers $n_i$ in a
sum of two integers: $n_i=r_i+s_i$.

\item ({\it Commitment.})  A sends $s_1$ to B, and B  sends
$r_2$ to A.
 Then, A sends $r_1+r_2$ to D, and B  sends $s_1+s_2$ to D.

\item ({\it Decommitment.}) D  reveals $r_1+r_2+s_1+s_2 =
n_1+n_2$ both to A   and B.

\item Now A knows $(n_1+n_2)-n_1 = n_2$, and B knows $(n_1+n_2)-n_2 =
n_1$, so cheating by either party is impossible.

\end{enumerate}

\section{$k$-$n$ oblivious transfer}
\label{oblivious}

An oblivious transfer protocol  is a protocol by which a sender
sends some information to the receiver, but remains oblivious as to
what is received. The first form of oblivious transfer was
introduced in 1981 by  Rabin \cite{Rabin}. Rabin's oblivious
transfer was later shown to be equivalent to ``1-2 oblivious
transfer"; the latter was subsequently generalized to $1$-$n$
oblivious transfer and to $k$-$n$ oblivious transfer \cite{BCR}. In
the latter case, the receiver obtains a set of $k$ messages from a
collection of $n$ messages. The set of $k$ messages may be received
simultaneously (``non-adaptively"), or they may be requested
consecutively, with each request based on previous messages
received. All the aforementioned constructions use encryption, so in
particular they use one-way functions. The first proposal that did
not use one-way functions (and therefore offered unconditionally
secure oblivious transfer) appeared in the paper by Rivest
\cite{Rivest} that we have already cited in our Section
\ref{commitment_2}.

In this section, we offer an  unconditionally secure $k$-$n$
oblivious transfer protocol that is essentially different from that
of Rivest in a similar way that our bit commitment protocol in
Section \ref{commitment_2} is different from Rivest's
unconditionally secure bit commitment protocol \cite{Rivest}. More
specifically, the extra participant in \cite{Rivest} is a ``trusted
initializer", which means, in particular, that  {\bf (i)} he is
allowed to generate randomness;   {\bf (ii)} he can ``consciously"
{\it transmit} information {\it to} ``real" participants over secure
channels.

By contrast, our extra participant is a ``dummy", i.e., {\bf (i)} he
is not allowed to generate randomness; {\bf (ii)}  he can  {\it
receive} information {\it from} ``real" participants over secure
channels, but he {\it  transmits} information upon specific requests
only.

Again,  let A (Alice) and B  (Bob) be two ``real" participants, and
D (Dummy) the ``dummy", e.g., a computer. Suppose A has a collection
of $n$ messages,  and B wants to obtain   $k$ of these messages,
without A knowing which   messages  B has received. Suppose that all
messages are integers $m_i, ~1 \le i \le n$.

\begin{enumerate}

\item  A randomly splits her  integers $m_i$ in a
sum of two integers: $m_i=r_i+s_i$.

\item A sends the (ordered)  set of all $r_i,  ~1 \le i \le n,$ to D, and the (ordered)
set of  all $s_i,  ~1 \le i \le n,$ to B.

\item B sends to D the set of indices  $j_1, \ldots, j_k$ corresponding to the
messages $m_j$ he wants to receive.

\item D sends to B the (ordered)  set $r_{j_1}, \ldots, r_{j_k}$.

\item B recovers $m_{j_1}, \ldots, m_{j_k}$ as a sum of relevant
$r_j$ and   $s_j$.

\end{enumerate}

\section{Mental poker}
\label{poker}

``Mental poker" is the common name for a set of cryptographic
problems that concerns playing a fair game over distance without the
need for a trusted third party. One of the ways to describe the
problem is: how can 2 players deal cards fairly over the phone?
Several protocols for doing this have been suggested, including
\cite{SRA}, \cite{Crepeau}, \cite{GM} and \cite{BF}. As with the bit
commitment, it is rather obvious that a fair card dealing to two
players over distance is impossible without a one-way function, or
even a one-way function with trapdoor. However,   it turns out to be
possible if the number of players is at least 3, assuming, of
course, that there are secure channels for communication between at
least some of the players. In our proposal, we will be using $k$
secure channels for $k \ge 3$ players $P_1, \ldots, P_k$, and these
$k$ channels will be arranged in a circuit: $P_1 \to P_2 \to \ldots
\to P_k \to P_1$.


To begin with, suppose there are 3 players: $P_1$, $P_2$, and $P_3$
and 3 secure channels: $P_1 \to P_2 \to P_3  \to P_1$.

The first protocol, Protocol 1 below, is for distributing {\it all}
integers from 1 to $m$  to the players in such a way that each
player gets about the same number of integers. (For example, if the
deck that we want to deal has 52 cards, then two players should get
17 integers each, and one player should get 18 integers.) In other
words, Protocol 1 allows one to randomly split a set of $m$ integers
into 3 disjoint sets.


The second protocol, Protocol 2, is for collectively generating
random integers modulo a given integer $M$. This very simple but
useful primitive can be used: {\bf (i)} for collectively generating,
uniformly randomly, a permutation from the group $S_m$. This will
allow us to assign cards from a  deck of $m$ cards to the $m$
integers distributed by Protocol 1; {\bf (ii)} introducing ``dummy"
players as well as for ``playing" after dealing cards.

\subsection{Protocol 1}
\label{Protocol1}

For notational convenience, we are assuming below that we have to
distribute integers from 1 to $r=3s$ to 3 players.

To begin with, all players agree on a parameter $N$, which is a
positive integer of a reasonable magnitude,  say, 10.

\begin{enumerate}

\item each player $P_i$ picks, uniformly randomly, an integer (a ``counter") $c_i$ between 1 and
$N$, and keeps it private.

\item $P_1$  starts with the ``extra"  integer  0  and sends it to $P_2$.

\item   $P_2$  sends  to $P_3$ either  the  integer $m$ he got from
$P_1$, or  $m+1$. More specifically, if  $P_2$ gets from $P_1$ the
same integer $m$ less than or equal to $c_2$ times, then he sends
$m$ to $P_3$; otherwise, he sends $m+1$ and keeps $m$ (i.e., in the
latter case $m$ becomes one of ``his" integers). Having sent out
$m+1$, he ``resets his counter", i.e., selects, uniformly randomly
between 1 and $N$, a new $c_2$. He also resets his counter if he
gets the number $m$ for the first time, even if he does not keep it.

\item $P_3$ sends  to  $P_1$ either  the  integer $m$ he got from
$P_2$, or  $m+1$. More specifically, if  $P_3$ gets from $P_2$ the
same integer $m$ less than or equal to  $c_3$ times, then he sends
$m$ to $P_1$; otherwise, he sends $m+1$ and keeps $m$. Having sent
out $m+1$, he selects a new counter $c_3$. He also resets his
counter if he gets the number $m$ for the first time, even if he
does not keep it.

\item   $P_1$  sends  to $P_2$ either  the  integer $m$ he got from
$P_3$, or  $m+1$. More specifically, if $P_1$ gets from $P_3$ the
same integer $m$ less than or equal to  $c_1$ times, then he sends
$m$ to $P_2$; otherwise, he sends $m+1$ and keeps $m$. Having sent
out $m+1$, he selects a new counter $c_1$. He also resets his
counter if he gets the number $m$ for the first time, even if he
does not keep it.

\item This procedure continues until one of the players gets $s$
integers (not counting the ``extra"  integer  0). After that, a
player who already has $s$ integers just ``passes along" any integer
that comes his way, while other players keep following the above
procedure until they, too, get $s$ integers.

\item The protocol ends as follows. When all $3s$ integers, between 1 and  $3s$, are
distributed, the player who got the last integer, $3s$, keeps this
fact to himself and passes this integer along as if he did not
``take" it.

\item The process ends when the integer $3s$ makes $N+1$ ``full
circles".

\end{enumerate}



\medskip

We note that the role of the ``extra"  integer  0 is to prevent
$P_3$ from knowing that $P_2$ has got the integer 1 if it happens so
that $c_2=1$ in the beginning.

We also note that this protocol can be generalized to   arbitrarily
many players in the  obvious way, if there are $k$ secure channels
for communication between  $k$ players, arranged in a circuit.

\subsection{Protocol 2}
\label{Protocol2}

Now we describe a protocol for  generating  random integers modulo
some integer $M$ collectively by 3 players. As in Protocol 1, we are
assuming that there are secure channels for communication between
the players, arranged in a circuit.


\begin{enumerate}

\item  $P_2$ and  $P_3$ uniformly randomly and independently select private
integers $n_2$ and $n_3$  (respectively) modulo $M$.

\item $P_2$ sends $n_2$ to $P_1$, and
$P_3$ sends $n_3$ to $P_1$.

%

\item $P_1$ computes the sum $m = n_2+n_3$ modulo $M$.


\end{enumerate}

Note that neither $P_2$ nor $P_3$ can cheat by trying to make a
``clever" selection of their $n_i$ because the sum, modulo $M$, of
any integer with an  integer uniformly distributed between 0 and
$M-1$, is an  integer uniformly distributed between 0 and $M-1$.

Finally, $P_1$ cannot cheat simply because he does not really get a
chance: if he miscalculates $n_2+n_3$ modulo $M$, this will be
revealed at the end of the game. (All players keep contemporaneous
records of all transactions, so that at the end of the game,
correctness could be verified.)

To generalize Protocol 2 to arbitrarily many players $P_1, \ldots,
P_k, ~k \ge 3,$ we can just engage 3  players at a time in running
the above protocol. If, at the same time, we want to keep the same
circular arrangement of secure channels between the  players that we
had in Protocol 1, i.e., $P_1 \to P_2 \to \ldots P_k \to P_1$, then
3 players would have to be $P_{i+1}$, $P_{i}$, $P_{i+2}$, where $i$
would run from 1 to $k$, and the indices are considered modulo $k$.

Protocol 2 can now be used to collectively generate, uniformly
randomly, a permutation from the group $S_m$. This will allow us to
assign cards from a  deck of $m$ cards to the  $m$ integers
distributed by Protocol 1. Generating a random permutation from
$S_m$ can be done by taking a random integer between 1 and $m$
(using Protocol 2) sequentially, ensuring that there is no
repetition. This ``brute-force" method will require occasional
retries whenever the random integer picked is a repeat of an integer
already selected. A simple algorithm to generate a permutation from
$S_m$ uniformly randomly  without retries, known as the {\it Knuth
shuffle}, is to start with the identity permutation or any other
permutation, and then go through the positions $1$ through $(m-1)$,
and for each position $i$ swap the element currently there with an
arbitrarily chosen element from positions $i$ through $m$, inclusive
(again, Protocol 2 can be used here to produce a random integer
between $i$ and $m$). It is easy to verify that any permutation of
$m$ elements will be produced by this algorithm with probability
exactly $\frac{1}{m!}$, thus yielding a uniform distribution over
all such permutations.


After this is done, we have $m$ cards distributed uniformly randomly
to the players, i.e., we have:

\begin{proposition}
If $m$ cards are distributed to $k$ players using Protocols 1 and 2,
then the probability for any particular card to be distributed to
any particular player is $\frac{1}{k}$.
\end{proposition}

\subsection{Using ``dummy" players while dealing cards}
\label{dummy}

We now show how a combination of Protocol 1 and  Protocol 2 can be
used to deal cards to just 2 players. If we have 2 players, they can
use a ``dummy" player (e.g. a computer), deal cards to 3 players as
in Protocol 1, and then just ignore the ``dummy"'s cards, i.e.,
``put his cards back in the deck". We note that the ``dummy" in this
scenario would not generate randomness; it will be generated for him
by the other two players using Protocol 2. Namely, if we call the
``dummy" $P_3$, then the player $P_1$ would randomly generate
$c_{31}$ between 1 and $N$ and send it to $P_3$, and $P_2$ would
randomly generate $c_{32}$ between 1 and $N$ and send it to $P_3$.
Then $P_3$ would compute  his random  number as $c_{3} =
c_{31}+c_{32}$ modulo $N$.

Similarly, ``dummy" players can help $k$ ``real" players each get a
fixed number $s$ of cards, because Protocol 1 alone is only good for
distributing {\it all} cards in the deck to the players, dealing
each player about the same number of cards. We can introduce $m$
``dummy" players so that $(m+k) \cdot s$ is approximately equal to
the number of cards in the deck, and position all the ``dummy"
players one after another as part of a circuit $P_1 \to P_2 \to
\ldots P_{m+k} \to P_1$. Then we use Protocol 1 to distribute all
cards in the deck to $(m+k)$ players taking care that each ``real"
player gets exactly $s$  cards. As in the previous paragraph,
``dummy" players have ``real" ones generate randomness for them
using Protocol 2.

After all cards in the deck are distributed to $(m+k)$ players,
``dummy" players send  all their cards to one of them; this ``dummy"
player now becomes a ``dummy dealer", i.e., he  will  give out
random cards from the deck to ``real" players as needed in the
course of a subsequent game, while randomness itself will be
supplied to him by ``real" players using Protocol 2.

\section{Summary of the properties of our card dealing (Protocols 1 and 2)}
\label{properties}

Here we summarize the properties of our Protocols 1 and 2 and
compare, where appropriate, our protocols to the card dealing
protocol of \cite{BF}.
\bigskip

\noindent {\bf 1. Uniqueness of cards.}  Yes, by the very design of
Protocol 1.

\medskip

\noindent {\bf 2. Uniform random distribution of cards.}  Yes,
because of  Protocol 2; see our Proposition 1 in Section
\ref{Protocol2}.

\medskip

\noindent {\bf 3. Complete confidentiality of cards.}  Yes, by the
design of Protocol 1.
\medskip

%
%

\noindent {\bf 4. Number of secure channels for communication
between $k \ge 3$ players:}  ~$k$, arranged in a circuit.

By comparison,  the card dealing protocol of \cite{BF} requires $3k$
secure channels.
\medskip

\noindent {\bf 5. Average number of transmissions between  $k \ge 3$
players:}  ~$O(\frac{N}{2}mk)$, where $m$ is the number of cards in
the deck, and $N \approx 10$. This is because in Protocol 1, the
number of circles (complete or incomplete) each integer makes is
either 1 or the minimum of all the counters $c_i$ at the moment when
this integer completes the first circle. Since the average of $c_i$
is at most $\frac{N}{2}$, we get the result because within one
circle (complete or incomplete) there are at most $k$ transmissions.
We note that in fact, there is a precise formula for the average of
the minimum of $c_i$ in this situation: ~$\frac{\sum_{j=1}^N
j^k}{N^k}$, which is less than $\frac{N}{2}$ if $k \ge 2$.

By comparison, in the protocol of \cite{BF} there are $O(mk^2)$
transmissions.
\medskip

\noindent {\bf 6. Total length of transmissions between  $k \ge 3$
players:}  ~$\frac{N}{2}mk \cdot \log_2 m$ ~bits. This is just the
average number of transmissions times the length of a single
transmission, which is a positive integer between 1 and $m$.

By comparison, total length of transmissions in \cite{BF} is $O(m
k^2 \log k)$.

\medskip

\noindent {\bf 7. Computational cost of Protocol 1:} ~0 (because
there are no computations, only transmissions).

By comparison,  the protocol of \cite{BF} requires computing
products of up to $k$ permutations from the group $S_k$ to deal just
one card; the total computational cost therefore is $O(m k^2 \log
k)$.

\section{Secret sharing}
\label{secret}

Secret sharing refers to method for distributing a secret amongst a
group of participants, each of whom is allocated a share of the
secret. The secret can be reconstructed only when a sufficient
number of shares are combined together; individual shares are of no
use on their own.

More formally, in a secret sharing scheme there is one dealer and
$k$ players. The dealer gives a secret to the players, but only when
specific conditions are fulfilled. The dealer accomplishes this by
giving each player a share in such a way that any group of $t$ (for
threshold) or more players can together reconstruct the secret but
no group of fewer than $t$ players can. Such a system is called a
$(t, k)$-threshold scheme (sometimes   written as a $(k,
t)$-threshold scheme).

Secret sharing was invented by   Shamir \cite{Shamir} and Blakley
\cite{Blakley}, independent of each other, in 1979. Both proposals
assumed secure channels for communication between the dealer and
each player. In our proposal here, the number of secure channels  is
equal to $2k$, where $k$ is the number of players, because in
addition to the secure channels between the dealer and each  player,
we have $k$ secure channels for communication between the players,
arranged in a circuit: $P_1 \to P_2 \to \ldots \to P_k \to P_1$.

The advantage over Shamir's and other known secret sharing schemes
that we are going to get here is that nobody, including the dealer,
ends up knowing  the shares (of the secret) owned by any particular
players. The disadvantage is that our scheme is a $(k, k)$-threshold
scheme only.

We start by describing a subroutine for distributing shares by the
players among themselves. More precisely, $k$  players want to split
a given number in a sum of  $k$ numbers, so that each  summand is
known to one player only, and each player knows one summand only.

\subsection{The Subroutine (distributing shares by the players among themselves)}

Suppose a player $P_i$ receives  a   number $M$ that has to be split
in a sum of  $k$ private numbers. In what follows, all indices are
considered modulo $k$.

\begin{enumerate}

\item $P_i$ initiates the process by sending  $M- m_i$ to $P_{i+1}$, where $m_i$
is a random number (could be positive or negative).

\item Each subsequent $P_j$   does the following. Upon
receiving a number $m$ from $P_{j-1}$, he subtracts a random number
$m_j$ from $m$ and sends the result to $P_{j+1}$. The number $m_j$
is now $P_j$'s secret summand.

\item  When this process gets back to $P_{i}$, he adds $m_i$ to
whatever he got from  $P_{i-1}$; the result is his secret summand.

\end{enumerate}

Now we get to the actual secret sharing protocol.

\subsection{The protocol (secret sharing $(k, k)$-threshold
scheme)}

The dealer $D$ wants to distribute shares  of a secret number $N$ to
$k$ players $P_i$ so that, if $P_i$ gets a  number $s_i$, then
~$\sum_{i=1}^k s_i = N$.

\begin{enumerate}

\item $D$ arbitrarily splits $N$ in a sum of $k$  integers: $N= \sum_{i=1}^k
n_i$.

\item The loop: at Step $i$ of the loop,  $D$ sends $n_i$ to $P_i$, and $P_i$
initiates the above Subroutine to distribute shares $n_{ij}$ of
$n_i$ among the players, so that ~$\sum_{j=1}^k n_{ij} = n_i$.

\item After all $k$ steps of the loop are completed, each  player
$P_i$ ends up with $k$ numbers $n_{ji}$ that sum up to $s_i=
\sum_{j=1}^k n_{ji}$. It is obvious that ~$\sum_{i=1}^k s_i = N$.

\end{enumerate}

\vskip .5cm

\noindent {\it Acknowledgement.} Both authors are grateful to  Max
Planck Institut f\"ur Mathematik, Bonn for its hospitality during
the work on this paper.


\baselineskip 11 pt


\begin{thebibliography}{ABC}



\bibitem{BF}
I. B\'ar\'any, Z. F\"uredi, {\it  Mental poker with three or more
players}, Inform. and Control {\bf 59} (1983),  84–-93.


\bibitem{Blakley}
G. R. Blakley,  {\it  Safeguarding cryptographic keys}, Proceedings
of the National Computer Conference {\bf  48} (1979), 313-–317.

\bibitem{BCR}
G. Brassard, C. Cr\'epeau and J.-M. Robert, {\it  All-or-nothing
disclosure of secrets}, In Advances in Cryptology -—  CRYPTO '86,
pp. 234–-238,   Lecture Notes Comp. Sc. {\bf 263}, Springer, 1986.


\bibitem{Chaum}
D. Chaum, C. Cr\'epeau, and I. Damgard,  {\it  Multiparty
unconditionally secure protocols (extended abstract)}, Proceedings
of the Twentieth ACM Symposium on the Theory of Computing, ACM,
1988, pp. 11-–19.

\bibitem{Crepeau}
C. Cr\'epeau, {\it A zero-knowledge poker protocol that achieves
confidentiality of the players' strategy or how to achieve an
electronic poker face},  Advances in cryptology -— CRYPTO '86, pp.
239–-247, Lecture Notes Comp. Sc.  {\bf  263}, Springer, 1986.

\bibitem{DGK}
I. Damgard, M.  Geisler, M. Kroigard,  {\it  Homomorphic encryption
and secure comparison},  Int. J. Appl. Cryptogr. {\bf   1} (2008),
22-–31.

\bibitem{DI}
I. Damgard, Y. Ishai,  {\it  Scalable secure multiparty
computation}, Advances in cryptology -— CRYPTO 2006, 501–-520,
Lecture Notes in Comput. Sci. {\bf  4117}, Springer, Berlin, 2006.

\bibitem{OG}
O. Goldreich,  {\em   Foundations of Cryptography: Volume 1, Basic
Tools}. Cambridge University Press, 2007.

\bibitem{GM}
S. Goldwasser and  S. Micali, {\it  Probabilistic Encryption and How
to Play Mental Poker Keeping Secret All Partial Information}, in
Proceedings of the 14th Annual ACM symp. on Theory of computing,
ACM-SIGACT, May 1982, pp. 365--377.

\bibitem{GM2}
S. Goldwasser, S. Micali, {\it   Probabilistic encryption}, J.
Comput. System Sci.  {\bf 28} (1984),   270–-299.



\bibitem{Grigoriev2}
D. Grigoriev, I. Ponomarenko, {\it Constructions in public-key
cryptography over matrix groups},  Contemp. Math., Amer. Math. Soc.
{\bf  418} (2006), 103--119.

\bibitem{Luby}
R. Impagliazzo  and  M. Luby, {\it One-way functions are essential
for complexity based cryptography}, in: FOCS'89, IEEE Computer
Society, 1989, pp. 230--235.

\bibitem{Menezes}
A. Menezes, P. van Oorschot, and S. Vanstone, {\sl  Handbook of
Applied Cryptography}, CRC-Press 1996.

\bibitem{Rabin}
M. Rabin, {\it  How to exchange secrets by oblivious transfer},
Technical Report TR-81, Aiken Computation Laboratory, Harvard
University, 1981.

\bibitem{Rivest}
R. Rivest, {\it Unconditionally Secure Commitment and Oblivious
Transfer Schemes Using Private Channels and a Trusted Initializer},
preprint, 1999.


\bibitem{Shamir}
A. Shamir,  {\it  How to share a secret}, Comm. ACM   {\bf 22}
(1979), 612-–613.

\bibitem{SRA}
A. Shamir, R. Rivest, and L. Adleman, {\it Mental poker}, Technical
Report LCS/TR-125, Massachusetts Institute of Technology, April
1979.


\bibitem{Yao}
A. C. Yao, {\it  Protocols for secure computations} (Extended
Abstract),  23rd annual symposium on foundations of computer science
(Chicago, Ill., 1982),  160--164, IEEE, New York, 1982.







\end{thebibliography}
\end{document}